\documentclass[12pt]{article}

\catcode`\@=11

\global\arraycolsep=2pt
\usepackage{amsbsy,amssymb,latexsym,amsfonts,amsmath}
\usepackage{graphicx,color}

\begin{document}
\begin{flushright}
\parbox{4.2cm}
{IPMU13-0186}
\end{flushright}

\vspace*{0.7cm}

\begin{center}
{ \Large Vector Beta function}
\vspace*{1.5cm}\\
{Yu Nakayama}
\end{center}
\vspace*{1.0cm}
\begin{center}
{\it Kavli Institute for the Physics and Mathematics of the Universe (WPI),  \\ Todai Institutes for Advanced Study,
University of Tokyo, \\ 
5-1-5 Kashiwanoha, Kashiwa, Chiba 277-8583, Japan}
\vspace{3.8cm}
\end{center}

\begin{abstract}
We propose various properties of renormalization group beta functions for vector operators in relativistic quantum field theories. We argue that they must satisfy compensated gauge invariance, orthogonality with respect to scalar beta functions, Higgs-like relation among anomalous dimensions and a gradient property. We further conjecture that non-renormalization holds if and only if the vector operator is conserved.
The local renormalization group analysis guarantees the first three within power counting renormalization. We verify all the conjectures in conformal perturbation theories and holography in the weakly coupled gravity regime.
\end{abstract}

\thispagestyle{empty} 

\setcounter{page}{0}

\newpage

\section{Introduction}
The introduction of the renormalization group beta function in quantum field theories was one of the major triumphs in theoretical physics. It has enabled us to understand the high or low energy universal behavior of the quantum field theories.  It has revealed the zero-charge problem in QED or $\phi^4$ theory in $d=1+3$ dimension, asymptotic freedom of QCD, possibilities of grand unification, the universality  of the critical phenomena and so on. 

In this paper, we study the renormalization group beta functions for vector operators. The background source for the vector operators breaks Poincar\'e invariance explicitly unlike the scalar coupling constant, but the importance of such symmetry breaking has been understood not only in condensed matter systems but also in high energy physics or cosmology. The introduction of temperature and chemical potential are nothing but such deformations, and our Freedman-Robertson universe has a preferred ``time" direction. The anomalous dimensions of vector operators will be important to understand the renormalization of the (partially conserved) vector operators such as vector mesons in QCD. The introduction of the background vector fields also play crucial roles in preserving supersymmetry in curved background in applications to twisting and localization. In all these cases, the vector beta functions govern the renormalization properties of such anisotropy.

While eventually one should compute the vector beta functions in each models either perturbatively or non-perturbatively, our approach in this paper focuses on the general properties. We claim various model independent properties of the vector beta functions. One may be able to check our proposals in concrete models such as our standard models. To show the validity of our general argument, we give the first principle derivation of some of the proposals based on the local renormalization group analysis.  In complement, we present the checks of all the proposed properties in conformal perturbation theories and holography. Note that conventional perturbation theories around Gaussian fixed point is just one example of conformal perturbation theories. The holographic dual may be regarded as the strongly coupled limit. 

One important tool to formulate and derive these properties of vector beta functions is the local renormalization group analysis \cite{Osborn:1991gm}\cite{JO}. We uplift all the ``coupling constants" into space-time dependent background fields. We will see that main source of the vector beta functions come from the space-time dependent ``coupling constants". 
The renormalization group flow of these space-time dependent coupling constants is governed by the local renormalization group transformation and the local renormalization group equation (a.k.a. local Callan-Symanzik equation). The consistency condition for the local renormalization group flow is non-trivial, which give certain integrability conditions on the beta functions. 

The local renormalization group together with large $N$ behavior of correlation functions of $d$-dimensional quantum field theories give birth to the holographic interpretation from the $d+1$ dimensional space-time, which was emphasized in \cite{Lee:2012xba}\cite{Lee:2013dln}. One of the properties of the vector beta functions we propose, i.e. the gradient property is motivated by the observation that the gradient property is crucial to remove the so-called scale reversal symmetry breaking term in the bulk, pushing it to the boundary contributions \cite{Lee:2013dln}. This is actually closely related to the integrability conditions on the local renormalization group which we mentioned above. We believe that the existence of such non-trivial properties of beta functions may explain the nature of the space-time realized in holography e.g. why we have space-time diffeomorphism or why time is different from space and so on.

The organization of the paper is as follows. We begin with the definition and the properties of the vector beta functions in section 2. We give arguments for the proposed properties from local renormalization group analysis. Some of the claims  still remain conjectures. In section 3, we give checks of all the conjectures in conformal perturbation theories and holography. We explain the computation of the vector beta functions from holography in detail. In section 4, we give further discussions on the vector beta functions with possible applications.

\section{Properties}
The vector beta function is defined as a response of the vector field source
 in the effective action at energy scale $\mu$ with scalar coupling constants $g^I$ and the vector source $a^a_\mu$
\begin{align}
S_{\mathrm{eff}}[X,g^I, a_\mu; \mu] = S_0[X]+ \int d^dx \left(g^I O_I + a^a_\mu J_a^\mu + O(a_\mu^2)  \right)
\end{align}
under the renormalization:
\begin{align}
\frac{da^a_{\mu}} {d\log \mu} = \beta^a_\mu \ , \label{vectorbeta}
\end{align}
which is analogous to the scalar beta functions
\begin{align}
\frac{dg^{I}}{d\log \mu} = \beta^I \ . 
\end{align}
Here the notation might suggest $X$ are fundamental dynamical fields and $O^I$ and $J^a_\mu$ are composite operators that can be constructed out of $X$, but such microscopic Lagrangian formulation may not be necessary in the following argument.
We also note that the definition of the effective action beyond the linear order in $a^a_\mu$ and other coupling constants $g^I$ is non-universal e.g. by seagull terms, and it can be 
scheme dependent in relation to contact term ambiguities. As in the scalar beta functions, these scheme dependence will affect the apparent value of the vector beta functions, but the physical predictions and the certain structures should remain invariant. This is nothing but the philosophy of the renormalization.

In order to understand the properties of the vector beta functions $\beta^a_\mu$ better, we would like to introduce more structures. First of all, we place the theory on a curved space-time with generic metric $g_{\mu\nu}(x)$. Secondly, we make source fields, such as vector source $a^a_\mu$ as well as scalar source $g^I$, space-time dependent (i.e. the source terms now look like $\int d^d x \sqrt{g(x)} \left( g^I(x) O_I(x) + a^a_\mu(x) J_a^\mu (x) + \cdots \right) $ in the effective action). There is some arbitrariness in this procedure at the non-linear level and/or higher derivative level, which is again related to the scheme dependence. When necessary we discuss the scheme dependence and ambiguities below.\footnote{A systematic classification of scheme dependence and anomaly in local renormalization group analysis in $d=3$ was done in \cite{Nakayama:2013wda}. For $d=2$ and $d=4$, we refer to \cite{Osborn:1991gm}\cite{JO}.} Once the scheme is fixed, one may define the operator insertion from Schwinger's quantum action principle
\begin{align}
\langle O_I(x) \rangle = -\frac{\delta W}{\delta g^I(x)} \ \cr
\langle J^\mu_a(x) \rangle = -\frac{\delta W}{\delta a^a_\mu(x)} \cr
\langle T_{\mu\nu}(x) \rangle = -2\frac{\delta W}{\delta g^{\mu\nu}(x)} \ , \label{actionp}
\end{align}
where, in Lagrangian field theories, Schwinger's source functional can be represented as the path integral form:
\begin{align}
e^{W} = \int \mathcal{D} X e^{-S_{\mathrm{eff}}[X,g^I,a_\mu,g_{\mu\nu} ;\mu]} \ . \label{sourcef}
\end{align}
Again, as long as Schwinger's quantum action principle holds, the expression \eqref{sourcef} is not necessary.
At this point, we note that we work in Euclidean field theories for definiteness but our argument should apply in Lorentzian signature as well. The path integral over $X$ and Schwinger's source functional $W$ should be regarded as a renormalized one, and \eqref{actionp} always gives a finite result. In the actual renormalization procedure, we may need more counter-terms than those necessary in conventional renormalization in flat space-time with constant coupling constants \cite{Jack:1990eb}, one of which is nothing but the vector beta functions as we will see below.
If we want to compute the multiple insertions of operators, we do have to be careful about the contact terms, which we will discuss later in relation to the anomalous dimensions.

Once the metric as well as source fields are space-time dependent, it is more natural to consider the space-time dependent renormalization group transformation rather than the space-time independent change of scale. Such space-time dependent renormalization has a close connection to the Weyl transformation of the metric $g_{\mu\nu}(x) \to e^{2\sigma(x)} g_{\mu\nu}(x)$ generalizing the rigid dilatation transformation with constant $\sigma$.

The response to the Weyl transformation is summarized in the trace of the energy-momentum tensor as 
\begin{align}
T^{\mu}_{\ \mu} = \beta^I O_I + \beta_{\mu}^a J_a^\mu + D_\mu (v^a J_a^\mu) + A_{\mathrm{anomaly}} \ . \label{traceidentity}
\end{align}
Here $A_{\mathrm{anomaly}}$ is the trace anomaly that only depends on the source fields.\footnote{The form of the anomaly term is not important for the most of our studies. One may find the detailed form in \cite{Osborn:1991gm}\cite{JO} for $d=2$ and $d=4$, in \cite{Nakayama:2013wda} for $d=3$ and in \cite{Grinstein:2013cka} for $d=6$.}
Or equivalently, from Schwinger's action principle, the trace identity \eqref{traceidentity} is equivalent to the claim that Schwinger's source functional $W$ in \eqref{sourcef} is annihilated by
the local renormalization group operator
\begin{align}
\Delta_{\sigma} = \int d^d x \sqrt{g} \left(2\sigma g_{\mu\nu} \frac{\delta}{\delta g_{\mu\nu}} + \sigma \beta^I \frac{\delta}{\delta g^I} + (\beta_\mu^a - (\partial_\mu \sigma)v^a ) \frac{\delta}{\delta a_\mu^a} \right) \  \label{localrgo} 
\end{align}
up to the Weyl anomaly given by a local functional of the source fields.

One may note that under the non-constant Weyl transformation, the renormalization of the vector source may contain the additional term that is proportional to the derivative of the Weyl generator or the space-time dependent renormalization scale as can be seen from the term $D_\mu (v^a J_a^\mu)$ in the trace identity \eqref{traceidentity} or the last term in \eqref{localrgo}. This did not appear in \eqref{vectorbeta} since for the constant scale transformation, it just gives a total derivative contribution to the renormalized action. Such a contribution to the trace of the energy-momentum tensor is known as the virial current and it plays a central role in the distinction between scale invariance and conformal invariance (see \cite{Nakayama:2013is} and references therein).\footnote{This is because such contributions are not affected by the constant scale transformation, but they may jeopardize the special conformal transformation since the integration by part is not possible.}

Before going on, while our definitions of the beta functions are applicable in Wilsonian renormalization group and we believe the properties we will discuss in this paper are more or less true there with minor modifications, we work in the power-counting renormalization scheme, where $g^I$ are ``dimensionless" or ``marginal" coupling constants.
In practice, our discussion covers most of the situations when beta functions are meaningfully calculable. 
They contain any perturbation theories, weakly coupled holography, and supersymmetric field theories. After all the usefulness of the philosophy of the Wilsonian renormalization group is that we can only keep the marginal or relevant operators in the far infrared to deduce the universal structure of the systems.
The inclusion of relevant deformations such as mass operators are possible along the line of the discussions \cite{Osborn:1991gm}\cite{JO}\cite{Nakayama:2013wda} with certain amount of additional complexity that will not affect the most of the discussions and we will mostly suppress. At one point later in section 2.2, however, we will discuss the possible mixing of vector operators with derivative of other relevant operators because our interest is how the vector operator evolves under the renormalization group flow and it is important to learn that there exists a scheme where we can tame such mixing.

\subsection{Conjectures}
The properties of the scalar beta functions have been much studied in the literature. One of the most important one would be the generalized gradient flow. It is conjectured \cite{Wallace:1974dy}\cite{Zamolodchikov:1986gt} that the scalar beta functions are generated from the potential function $\hat{a}(g)$ as 
\begin{align}
\beta^I = \chi^{IJ} \frac{\partial \hat{a}(g)}{\partial g^J} \ . \label{gradient}
\end{align}
It was originally understood that $\chi^{IJ}$, which may depend on the coupling constant $g^I$, is positive definite symmetric matrix interpreted as a ``metric" on the coupling constant space, but the more general argument allows non-symmetric part \cite{Osborn:1991gm}\cite{JO}. Moreover, we would be careful about the meaning of the scalar beta functions when the theory contains vector operators and  the beta functions can be ambiguous, 
which we will come back in a moment.\footnote{More precisely, the gradient formula works for the gauge invariant $B^I$ function which we will define around \eqref{Bfunction}.} 

To understand the intrinsic physics, we note that in $d=2$ and $d=4$, it was argued that $\hat{a}(g)$ is related to the coefficient of the Euler term in the Weyl anomaly when we evaluate it at the conformal fixed point. As a consequence, the generalized ``$c$-theorem" \cite{Zamolodchikov:1986gt}\cite{Komargodski:2011vj}  follows from \eqref{gradient} by using the renormalization group invariance of Schwinger's source functional $\frac{d}{d\log\mu}W = \beta^I \partial_I W$ for the flow induced by the scalar coupling constants. 
The gradient formula \eqref{gradient} together with the claim that $\hat{a}(g)$ does not depend explicitly on $\mu$ implies
\begin{align}
\frac{d\hat{a}(g(\mu))}{d\log\mu} = \chi_{IJ} \beta^I \beta^J \ 
\end{align}
under such a flow.
In particular, when $\chi_{IJ}$ is positive definite, the flow is monotonic.

Our main focus of this paper is the properties of the vector beta functions. 
We claim that the vector beta functions satisfy the following properties:
(i) compensated gauge invariance, (ii) orthogonality with scalar beta functions (iii) Higgs-like relations among anomalous dimensions, (iv) gradient property and (v) non-renormalization for the conserved current. To make the statement unambiguous, we need to fix the renormalization group scheme.

(i) Compensated gauge invariance

The renormalization group invariance of Schwinger's source functional means that generically, the vector beta functions must be expanded as 
\begin{align}
\beta^a_\mu = \rho^a_I(a_\mu, g) \partial_\mu g^I + \lambda^{a}_{b}(g,a_\mu) a^{b}_{\mu} \ . \label{generalf}
\end{align}
If we assume the power-counting renormalization scheme, $\rho^a_I$ and $\lambda^{a}_b$ cannot depend on either derivatives of $g^I$ or the vector source $a^a_\mu$. 

In order to state the compensated gauge invariance of the vector beta functions, we assume that the coupling constants $g^I$ form certain representations of the compensated symmetry group $\mathcal{G}$ generated by $J^\mu_a$. 
The broken conservation means that we assume the operator identity
\begin{align}
D_\mu J^\mu_a = h_{ab} T^{bI}_{\ \ J} g^J O_I \ . \label{identity}
\end{align}
As a consequence, Schwinger's source functional must be invariant under the compensated gauge transformation
\begin{align}
g^I &\to g^I + (h_{ab} \omega^a T^{bI}_{\ \ J} g^J) \cr
a_\mu^a &\to a_\mu^a + D_\mu \omega^a \ \label{gauget}
\end{align}
for any Lie algebra element $w^a \in \mathfrak{g}$. Here, $D_\mu \omega^a = \partial_\mu \omega^a + a_\mu^b f^a_{bc} \omega^c $ is the covariant derivative. 
When non-zero, $h_{ab}$ can be set to be $\delta_{ab}$ by rescaling the source fields. We therefore regard $a_\mu^a$ as a connection of the compensated gauge group $\mathcal{G}$.

Going back to the vector beta functions, the compensated gauge invariance (together with the power-counting) requires that the two terms in \eqref{generalf} are combined into the covariant derivative
\begin{align}
\beta^a_\mu = \rho^a_I(g) D_\mu g^I \ 
\end{align}
where $g^I$ transforms as a certain representation of the group action $\mathcal{G}$, and $D_\mu g^I = \partial_\mu g^I + a^a_\mu h_{ab} T^{bI}_{\ \ J} g^J$ is the corresponding covariant derivative.

Due to the compensated gauge invariance, the meaning of ``coupling constants" and therefore beta functions can be ambiguous. In order to state the properties of the vector beta functions without such ambiguities, we will fix the gauge by demanding that there is no virial current contribution to the trace identity:
\begin{align}
T^{\mu}_{\ \mu} &= \beta^I O_I + \beta_{\mu}^a J_a^\mu + D_\mu (v^aJ^\mu_a) + A_{\mathrm{anomaly}} \cr
&= B^I O_I + B_{\mu}^a J_a^\mu  + A_{\mathrm{anomaly}} \ ,
\end{align}
where the scheme invariant quantities $B^I$ and $B^a_\mu$ are defined as
\begin{align}
B^I &= \beta^I + (h_{ab} v^a T^{b I}_{\ \ J} g^J) \cr
B^a_\mu &= \hat{\rho}_I^a D_\mu g^I = (\rho_I^a + \partial_I v^a) D_\mu g^I \ . \label{Bfunction}
\end{align}
For the consistency of \eqref{identity}, gauge transformation and local renormalization must commute \cite{JO} so that $B^I$ and $B^a_\mu$ should transform covariantly under the gauge transformation \eqref{gauget}.
The renormalization group scheme for the gauge dependence is thus fixed, and we hereafter assume that the renormalization of scalar and vector sources are defined with respect to $B^I$ and $B^a_\mu$ unless otherwise stated.
However we emphasize that if we compute the new functions $B^I$ and $B^a_\mu$ from the right hand side of \eqref{Bfunction} from $\beta^I$, $\beta^a_\mu$ and $v^a$ given in any gauge, it results in the gauge invariant quantities, and the physical predictions such as the existence of the conformal fixed point do not depend on the gauge we choose.

(ii) Orthogonality

Vector beta functions are orthogonal to scalar beta functions
\begin{align}
\hat{\rho}_I^a B^ I = 0 \ .
\end{align}
When applied in perturbation theories, this constraint should hold order by order in perturbation series, which gives an infinite set of relations among Feynman diagrams with number of different loops and topology.

(iii) Higgs-like relation between anomalous dimensions.

The anomalous dimension of the scalar operator $O^I$ is determined from the scalar beta functions as well as vector beta functions as 
\begin{align}
-\gamma^I_J = \partial_J B^I + (h_{ab} \hat{\rho}_J^a  T^{bI}_{\ \ K} g^K) \ .
\end{align}
More precisely, when the coupling constant is position dependent, there is a further mixing to the vector operators
\begin{align}
\frac{d}{d\log\mu} O_I = \gamma^{J}_I O_J - (\partial_I \hat{\rho}_J^a) D_\mu g^J J^\mu_a \ . 
\end{align}

Similarly, the anomalous dimension of the vector operator $J^\mu_a$ is given by
\begin{align}
-\gamma^{a}_{b} = \rho_I^a h_{bc} T^{cI}_{ \ \ J} g^J \ .
\end{align}
There is no additional mixing for the vector operator when the scalar source is position dependent
\begin{align}
\frac{d}{d\log\mu} J_a^\mu = \gamma_{a}^{b} J_b^\mu \ .
\end{align}

In these expressions, we have assumed that mixing to lower dimensional operators are absent. Possible additional mixing of the scalar operator to the second derivative of relevant operators e.g. $D^\mu D_\mu O_{(M)}$ will be discussed in section 2.2. The  mixing of the vector operator to the derivative of relevant operators can be removed by a scheme choice as we will also discuss in section 2.2.


(iv) Gradient property

Vector beta functions are generated by a (generalized) gradient flow from the gauge invariant local functional $S_{\mathrm{vector}}[a_\mu; g^I]$ as 
\begin{align}
B_\mu^a = \hat{\rho}_I^a D_\mu g^I = H^{ab} g_{\mu\nu} \frac{\delta S_{\mathrm{vector}}[a_{\mu};g^I]} {\delta a^b_\nu} \ ,
\end{align}
where $H^{ab}$ in general may contain anti-symmetric part. This is the analogue of the gradient flow of the scalar beta functions quoted at the beginning of this section.

(v) Non-renormalization when $J^a_\mu$ is conserved.

Vector beta functions vanish if and only if the corresponding current $J_\mu^a$ is conserved. From the property (iii), it is equivalent to vanishing of anomalous dimensions for $J_\mu^a$.

\subsection{Argument from local renormalization group} 
We are able to derive some of the properties of the vector beta functions proposed in the last section within the power-counting renormalization scheme from the local renormalization group analysis. We will show that they are a consequence of the local renormalization group equation and the integrability condition. This is similar to the situation for the scalar beta functions, whose properties such as the gradient flow can be derived from the local renormalization group analysis within the power-counting renormalization scheme in even space-time dimensions.\footnote{For scalar beta functions, it seems likely that the same properties should be true also in odd dimensions. We have not found any counterexamples as far as the author is aware of. We do not make any particular mentioning of dimensionality for vector beta functions.}

We begin with Schwinger's source functional. We assume the power-counting renormalizability so that the renormalized Schwinger's source functional is annihilated by the local renormalization operator
\begin{align}
\Delta_{\sigma} &= \int d^dx \sqrt{g} \left( 2\sigma g_{\mu\nu} \frac{\delta}{\delta g_{\mu\nu}} + \sigma \beta^I \frac{\delta}{\delta g^I}  \right. + \left. \left( \sigma \rho_I^a D_{\mu} g^I - (\partial_\mu \sigma) v^a \right) \frac{\delta}{\delta a^a_{\mu}} \right) \ . 
\end{align}
up to the Weyl anomaly if exists, which is given by a local functional constructed out of the source functions. 
We may introduce the dimensionful coupling constants here, but the following discussions will not change very much. If necessary, we will make a comment on the effect of dimensionful coupling constants below.
The invariance of Schwinger's source functional under the local renormalization group (up to anomaly) corresponds to the trace identity
\begin{align}
T^{\mu}_{\ \mu} = \beta^I O_I + (\rho_I^a D_\mu g^I)J^\mu_a + D_\mu (v^a J_a^\mu) + A_\mathrm{anomaly} \ .  \label{traceiden}
\end{align}

Let us consider the property (i) gauge invariance. This essentially follows from the definition of Schwinger's source functional because the operator identity that characterizes the non-conservation of the current operator 
\begin{align}
D_\mu J^\mu_a = h_{ab} T^{bI}_{\ \ J} g^J O_I 
\end{align}
means Schwinger's source functional must be invariant under the compensated gauge transformation\footnote{A part of the anomaly effects can be encoded here. For instance, the axial anomaly can be compensated by shifting the $\theta$ angle as a part of the transformation \eqref{gaugetransform}. In such cases, the covariant derivative is given by the Stuekelberg form $\partial_\mu \theta + A_\mu$. Similar situations occur when the current non-conservation is not directly related to the introduction of the coupling constant (e.g. higher dimensional irrelevant vector operators).}
\begin{align}
g^I &\to g^I + (h_{ab} \omega^a T^{bI}_{\ \ J} g^J) \cr
a_\mu^a &\to a_\mu^a + D_\mu \omega^a \ . \label{gaugetransform}
\end{align}
If the divergence of the vector operator $J_a^\mu$ is additionally given by the relevant operators e.g. $M^i O^{(M)}_i$, one may simply add further gauge transformations of $M^i$.
\begin{align}
M^i \to M^i + (h_{ab}\omega^a T^{b i}_{\ \ j} M^j) \ ,
\end{align}
where again we assume $M^i$ forms a certain representation of $\mathcal{G}$ with the representation matrix $T^{ai}_{\ \ j}$.

Since the energy-momentum tensor is gauge invariant, the vector beta functions must be constructed out of the covariant derivatives of the coupling constant from the assumption of the power-counting. The only available possibility is
\begin{align}
B_\mu^a = \hat{\rho}^a_I D_\mu g^I \ . \label{vectorbetaff}
\end{align}
This is the proposed form for the vector beta functions. 

In order to derive (ii) the orthogonality condition, we need to discuss the integrability condition of the local renormalization group operator on Schwinger's source functional.
The local renormalization group transformation is Abelian, so the integrability condition demands
\begin{align}
[\Delta_\sigma, \Delta_{\tilde{\sigma}}] = 0 \ .
\end{align}
This is known as Class 1 consistency condition in the classification of the consistency conditions from the local renormalization group analysis in \cite{Nakayama:2013wda}.\footnote{Let us recapitulate the classification. Class 1 consistency condition is the requirement of the integrability of the local renormalization group operator itself and it has no essential dependence on the space-time dimensionality $d$. Class 2 consistency condition is the condition on the anomaly and the precise form depends on $d$. }
For this to be satisfied, we have to demand \cite{Osborn:1991gm}
\begin{align}
 \int d^dx\sqrt{g} (\sigma \partial_\mu \tilde{\sigma} - \tilde{\sigma} \partial_\mu \sigma) B^I \hat{\rho}^a_I \frac{\delta}{\delta a^a_{\mu}} = 0 \ ,
\end{align}
or
\begin{align}
B^I \hat{\rho}_I^a = 0 \ , \label{rhocons}
\end{align}
This is the derivation of the orthogonality properties of the vector beta functions. The property played an essential role in deriving the gradient flow of the scalar beta functions in even dimensions \cite{JO}, but this orthogonality itself is true in any space-time dimensions.

To obtain (ii) the relations with anomalous dimension, we first derive the Callan-Symanzik equation for the correlation functions by further deriving Schwinger's source functional with respect to $g^I(x)$ and $a^a_\mu(x)$. After setting $D_\mu g^I =0$ with $a_\mu = 0$ and integrating over the space-time to get rid of one delta function, it gives 
\begin{align}
&\left(\frac{\partial}{\partial \log\mu} + B^I \frac{\partial}{\partial g^I}  \right) \langle O_{I_1}(x_1) O_{I_2}(x_2) \cdots J_{a_1}^\mu(y_1)J_{a_2}^\mu(y_2) \cdots \rangle \cr
 &=   \gamma^{J_1}_{I_1} \langle O_{J_1}(x_1) O_{I_2}(x_2) \cdots J_{a_1}^\mu(y_1) J_{a_2}^\mu(y_2)\cdots  \rangle +  \gamma^{J_2}_{I_2} \langle O_{I_1}(x_1) O_{J_2}(x_2) J_{a_1}^\mu(y_1) J_{a_2}^\mu(y_2) \cdots \rangle + \cdots \cr
& +  \gamma_{a_1}^{b_1} \langle O_{I_1}(x_1) O_{I_2}(x_2) \cdots J_{b_1}^\mu(y_1) J_{a_2}^\mu(y_2) \cdots  \rangle  +  \gamma_{a_2}^{b_2} \langle O_{I_1}(x_1) O_{I_2}(x_2) \cdots J_{a_1}^\mu(y_1) J_{b_2}^\mu(y_2) \cdots  \rangle + \cdots \cr
& + \text{contact terms} \ . 
\end{align}
up to contact terms with extra delta functions.  Here, the anomalous dimension matrix for the scalar operator is given by
\begin{align}
-\gamma^I_J = \partial_J B^I + (h_{ab} \hat{\rho}_J^a  T^{bI}_{\ \ K} g^K) \ , \label{anomalouss}
\end{align}
whose origin can be seen by applying $\frac{\delta}{\delta g^I}$ to the local renormalization group operator. In particular, we note that the second term comes from the vector beta functions.
Similarly the anomalous dimension matrix for the vector operator is given by 
\begin{align}
-\gamma^{a}_{b} = \rho_I^a h_{bc} T^{cI}_{ \ \ J} g^J \ , \label{anovect}
\end{align}
whose origin can be seen by applying $\frac{\delta}{\delta a_\mu^a}$ to the local renormalization group operator. 

We have called it the Higgs-like relation, which will be more manifest in the holographic construction. At the conformal fixed point, what is happening here is the branching of a spin one long representation into a spin one short representation (conserved current) and a spin zero long representation (scalar with dimension $d$).

One may actually keep the coupling constant space-time dependent in the above variation, and the extra variation means that the complete renormalization of the operator should be 
\begin{align}
\frac{d}{d\log\mu} O_I &= \gamma^{J}_I O_J - (\partial_I \hat{\rho}_J^a) D^\mu g^J J^\mu_a \cr
\frac{d}{d\log\mu} J^\mu_a &= \gamma^{b}_{a} J^\mu_b \cr
\frac{d}{d\log\mu} T^{\mu}_{\ \mu} &= 0
\end{align}
The last equality follows from the first two with the use of orthogonality property $\hat{\rho}^a_I B^I = 0$, which
means that the trace identity \eqref{traceidentity} is not renormalized.\footnote{When the theory contains relevant operators with dimension $d-2$, the trace identity can be renormalized.}

It is worthwhile commenting on the covariance of the anomalous dimension matrix $\gamma^{I}_J$ under the reparameterization of the coupling constants $\tilde{g}^J = \tilde{g}^J(g)$. Our formula is not covariant because the derivative appearing in \eqref{anomalouss} is ordinary derivative rather than covariant derivative. Of course without further information such as operator product expansion or trace anomaly, we have no candidate of the metric and hence connection at a generic point along the renormalization group flow. The consistency of the local renormalization group does not necessarily require that the anomalous dimensions defined here must transform covariantly under the reparameterization. One important point we should check however is that at the conformal fixed point where $B^I = 0$, the anomalous dimension matrix does not depend on the reparameterization of the coupling constants or renormalization scheme. This holds in our formula.

We would like to also mention possible mixing of the vector operators and the derivative of the scalar operators under the renormalization group flow when relevant operators exist.
The mixing is generated by the additional contribution to the local renormalization group operator with dimension close to $d-2$:
\begin{align}
\Delta_{\sigma,M} &= -\int d^dx \sqrt{g}\left(\sigma(2-\gamma^i_{j(M)}) M^j + \frac{1}{2(d-1)} \sigma R \eta^i + \sigma \delta^i_I (D^2 g^I) + \sigma \epsilon^i_{IJ} (D^\mu g^I D_\mu g^J) \right. \ \cr
 & \left. \left. + 2\partial_\mu \sigma (\theta^i_I D^\mu g^I) + (D^2 \sigma) \tau^i  \right) \frac{\delta}{\delta M^i} \right) \ . \label{mderiv}
\end{align}
By studying the local Callan-Symanzik equation, even in the background $D^\mu g^I =0$, there is a further renormalization of the operators by
\begin{align}
\frac{d}{d\log\mu} O_I &= \gamma^{J}_I O_J + \delta_{I}^i D^2 O_{i}^{(M)}  \cr
\frac{d}{d\log\mu} J_\mu^a &= \gamma^{a}_{b} J_\mu^a + 2\theta_I^i (T^{aI}_{\ \ J} g^J)D_\mu O_{i}^{(M)}  \cr
\frac{d}{d\log \mu}T^{\mu}_{\ \mu} & = \eta^i \Box O_{i}^{(M)} \ .
\end{align}
However, one may use the ambiguity in the renormalization group so that we may make some of the mixing vanish. Such ambiguity was called Class 2 ambiguity (scheme ambiguity) of the local renormalization group in \cite{Nakayama:2013wda}.\footnote{Let us again recapitulate the classification. Class 1 ambiguity is the compensated gauge ambiguity discussed above. Class 2 ambiguity is the ambiguity of the generalized reparameterization of the source function in Schwinger functional. Class 3 ambiguity is the ambiguity of the local counterterms.} In addition, for these equations to be consistent with the trace identity, the coefficients satisfy the integrability conditions which may be found in \cite{Osborn:1991gm}\cite{JO}\cite{Nakayama:2013wda} that we will not use in this paper.

Class 2 ambiguity of the local renormalization group was induced by the variation
\begin{align}
\delta \Delta_{\sigma} = [\mathcal{D},\Delta_{\sigma}] \ ,
\end{align}
where $\mathcal{D}$ is any local functional differential operator.
For our purpose, we consider the mixing between $R$, $D^2 g^I$ and $D_\mu g^I D^\mu g^J$ given by
\begin{align}
\mathcal{D} = \int d^dx \sqrt{g} \left(\frac{1}{2(d-1)}R h^i + (D^2 g^I) d^i_I + (D_\mu g^I D^\mu g^J) e^i_{IJ} \right) \frac{\delta}{\delta M^i} \ .
\end{align}
Under this scheme change associated with the field redefinition, we obtain
\begin{align}
\delta \eta^i &= (B^I \partial_I h^i - \gamma^i_{j(M)} h^j) \cr
\delta \tau^i &= -h^i + d^i_I B^I \cr
\delta \theta^i_I &= \frac{d-2}{2}d^i_I + \left(\partial_IB^J + \frac{1}{2}(\hat{\rho}_Ig)^J \right)d^i_J + e^i_{IJ} B^J \cr
\delta \delta_I^i &= (\tilde{\mathcal{L}}_{B,\hat{\rho}} -\gamma_{(M)}) d^i_I \cr
\delta \epsilon^i_{IJ} &= (\tilde{\mathcal{L}}_{B\,\hat{\rho}} - \gamma_{(M)}) e^i_{IJ} +
(\partial_I \partial_J B^K + (\partial_{(I}(\hat{\rho}_{J)})g)^K)d^i_K + 2d^i_K(\hat{\rho}_{(I})^K_{J)} \ ,
\end{align}
where $\tilde{\mathcal{L}}_{B,\hat{\rho}} X_I = \mathcal{L}_B X_I  + (\hat{\rho}_I^a T^{aJ}_{\ \ L}g^L)X_J$ and so on for higher tensors with $\mathcal{L}_B$ being the Lie derivative for vector $B^I$. By using the freedom associated with $d^i_I$, we can always choose $\theta^i_I = 0$ in order to remove the mixing of the vector operators with the derivative of scalar operators. In our discussions, we always implicitly assume $\theta^i_I = 0$. Sometimes it is convenient to choose $\tau^i = 0$ by adjusting $h^i$ so that the energy-momentum tensor becomes traceless when $B^I = 0$. However, this  choice does not guarantee that the energy-momentum tensor is not renormalized. For the non-renormalization of the energy-momentum tensor, we should require $\eta^i = 0$ which may or may not be possible.

The conditions so far obtained do not immediately lead to (iv) the gradient property, so it still remains a conjecture. Within the power-counting renormalization scheme, the most general candidate for the potential functional, which must be invariant under the compensated gauge transformation, is
\begin{align}
S_{\mathrm{vector}}[a_\mu;g^I] = \int d^d x \sqrt{g} G_{IJ} D_\mu g^I D^\mu g^J \ . 
\end{align}
In order for the vector beta functions to be given by the generalized gradient flow that we propose:
\begin{align}
B_\mu^a = \hat{\rho}_I^a D_\mu g^I = H^{ab} g_{\mu\nu} \frac{\delta S_{\mathrm{vector}}[a_{\mu};g^I]} {\delta a^b_\nu} \ ,
\end{align}
 they must satisfy
\begin{align}
\hat{\rho}_I^a D_\mu g^I = H^{ab} G_{IJ} h_{bc} (T^{cI}_{\ \ K} g^K) D_\mu g^J \ . \label{vectgrad}
\end{align}
We note this is a non-trivial constraint, and it is not invariant under the gauge transformation \eqref{gaugetransform} introduced above. To check the equality  \eqref{vectgrad} from the explicit computation of the vector beta functions at higher order, it is important to use the gauge invariant $B^a_\mu$ function.

Finally, let us discuss (v) the non-renormalization property. If one accepts the gradient formula realized in the power-counting renormalization scheme and the compensated gauge invariance as the violation of the conservation law, one can convince that when a current is conserved, then there exists a corresponding zero direction in the vector beta functions because certain combinations of \eqref{vectgrad} must vanish. 

Conversely, if we assume that the matrix $H^{ab}$ and $G_{IJ}$ are non-degenerate and do not contain any zero eigenvalues then when the vector beta function vanishes there exists a gauge transformation that does not act on $g^I$. Therefore there  should exist the corresponding vector operators that are conserved.

When the theory is unitary and conformal invariant, it is well-known that the anomalous dimension for the vector operator is zero if and only if it is conserved from the representation theory of the conformal algebra \cite{Mack:1975je}. This fact is completely in agreement with our claim. Our proposal can be regarded as a generalization beyond the conformal fixed point.

On the other hand, one immediate consequence of this non-renormalization of the vector operator is that scale invariance should imply conformal invariance. Assuming there is no mixing of the energy-momentum tensor to the other lower dimension operators under renormalization, the energy-momentum tensor must possess no anomalous dimension. The trace identity demands that the virial current $J_\mu$ in $ T^{\mu}_{\ \mu} = \partial^\mu J_\mu$ is not renormalized. However, the claim here is that without anomalous dimension, $J_\mu$ must be conserved and then the trace of the energy-momentum tensor actually vanishes.

Historically, the relation between scale invariance and conformal invariance was studied from the viewpoint of the scalar beta functions under the renormalization group flow with respect to $B^I$ functions. We may argue (e.g. by the gradient property) that the scale invariant fixed point, all of the scalar beta functions $B^I$ must vanish, and the theory at the fixed point must be conformal invariant \cite{Osborn:1991gm}. In a certain sense, the argument here is based on the complementary viewpoint from the vector beta functions. Recently, we have accumulated non-perturbative evidence for the conformal invariance at almost every (probably all under some assumptions) unitary scale invariant fixed points in $d=4$ \cite{Nakayama:2012nd}\cite{Luty:2012ww}\cite{Fortin:2012hn}\cite{Dymarsky:2013pqa}\cite{Farnsworth:2013osa}, and it would be interesting to see if the non-renormalization properties we proposed here can be shown along the similar line of reasoning.

\section{Checks}
In this section we check our proposals on the properties of vector beta functions in conformal perturbation theories and holography. We also note that the computation of the vector beta functions for supersymmetric field theories have been reported in \cite{Jack:1999aj}\cite{Nakayama:2012nd}\cite{Fortin:2012hc}\cite{JO} to all orders in perturbation theories, and we can explicitly  see that they satisfy our proposals.

\subsection{Conformal perturbation theory}
We can compute the vector beta functions explicitly within conformal perturbation theories. We begin with the operator product expansion
\begin{align}
O_I(x) O_J(y) = \frac{\delta_{IJ}}{(x-y)^{2d}} + \frac{\mathcal{C}_{IJK}}{(x-y)^d}O_K(y) + \frac{\mathcal{C}^{a}_{IJ}(x-y)_\mu}{(x-y)^{d+2}} J_{a}^\mu (y) + \cdots , \label{ope}
\end{align}
where the operator product expansion coefficient $\mathcal{C}_{IJK}$ is totally symmetric and $\mathcal{C}^a_{IJ} = -\mathcal{C}^a_{JI} = h_{ab} T_{IJ}^b$ is proportional to a certain representation matrix $T_{IJ}^a$ of the flavor symmetry group (denoted by $\mathcal{G}$ before) generated by $J_a^\mu$ at the conformal fixed point we expand around.\footnote{In this section, we raise and lower indices freely with respect to $\delta_{IJ}$ and $\delta_{ab}$ from the normalization at the conformal fixed point.} Indeed, the first order in conformal perturbation theory \cite{Friedan:2009ik}\cite{Nakayama:2013is}, the conservation of $J_a^\mu$ is broken by
\begin{align}
D_\mu J_a^\mu = \mathcal{C}^a_{IJ} g^J O_I \ .
\end{align}

At the second order in conformal perturbation theory, the vector beta functions can be computed as \cite{Nakayama:2013is}\cite{Nakayama:2013wda}\footnote{We suppress the numerical coefficients that depend on the space-time dimensionality $d$ that come from the sphere integral by appropriately normalizing the coupling constants.}
\begin{align}
\hat{\rho}^a_I = \mathcal{C}^a_{IJ} g^J = h_{ab} T_{IJ}^b g^J \ , \label{vectb}
\end{align}
We would like to check the various properties proposed in the last section. 

First of all, the gauge invariance follows naturally to the order we compute. Indeed, we could have used a different gauge by subtracting artificially more (up to total derivative) in the divergence that appears in the operator product expansions
\begin{align}
0 = \delta S_{\mathrm{eff}} = \log\mu \left(\int d^d x \sqrt{g} \left( g^I w_I^a D_\mu J^\mu_a \right)\right) - \log \mu \left( \int d^d x  \sqrt{g} \left( g^I w_I^a g^K \mathcal{C}_{LK}^a O_L \right) \right) \ ,
\end{align}
which leads to the gauge transformation
\begin{align}
\delta \beta_\mu^a &= w_I^a D_\mu g^I \cr
\delta \beta^I &= g^J w_J^a \mathcal{C}^a_{IK} g^K \ .
\end{align}
In this way, the conformal perturbation theories with redundant operators can be ambiguous \cite{Nakayama:2013is}.

However, typically there is no natural candidate for the gauge transformation parameter (e.g. $w^I_a$ in this above leading order example) constructed out of $g^I$ and the invariant tensor of the Lie algebra $\mathfrak{g}$ alone without further assuming the structure of the coupling constant space. This is a typical reason that at lower order in perturbation theories, we do not usually have to make clear distinctions between $\beta^I$, $\beta^a_\mu$ and $B^I$, $B^a_\mu$ .

To understand the orthogonality, we first realize that the scalar beta functions computed in the second order conformal perturbation theory satisfy the gradient property:
\begin{align}
B^I& = \frac{\partial C}{\partial g^I} \cr  
C &= \frac{\mathcal{C}_{IJK}}{3} g^I g^J g^K \ .
\end{align}
Up to the additive constant, we can identify $C$ with $\hat{a}(g)$ mentioned at the beginning of section 2.
To the order we compute, the metric $\chi^{IJ}$ in \eqref{generalf} is taken to be trivial.
Here, a crucial observation is that $C$ is invariant under the group action generated by $T^a_{IJ}$ since $\mathcal{C}_{IJK}$ is a symmetric invariant tensor which maps $R_I \otimes R_J \otimes R_K$ to $\mathbb{C}$. This directly means that the orthogonality condition must hold
\begin{align}
\hat{\rho}^a_I B^I &= h_{ab} T_{IJ}^b g^J \mathcal{C}_{IKL} g^K g^L \cr
&= h_{ab} T_{IJ}^b g^J \frac{\partial C}{\partial g^I} \cr
&= 0 \ .
\end{align}

One may compute the anomalous dimensions by using the formula in the last section and the data obtained from the conformal perturbation theories.
First of all, one may compute the anomalous dimension of the scalar operators $\gamma_{I}^J$ as
\begin{align}
-\gamma_{I}^J  = \partial_I B^J + (h_{ab} \hat{\rho}_J^a  T^{b}_{I K} g^K)   = 2\mathcal{C}_{IJK} g^K + O(g^2) \ .
\end{align}
As we discussed, there is an extra contribution that gives the mixing to the vector operator when the coupling constant is space-time dependent as
\begin{align}
\frac{d}{d\log\mu} O_I = \gamma_{I}^{J} O_J -
 h_{ab} T^b_{JI} D^\mu g^J J_\mu^a \ .
\end{align}
In a similar manner, we can compute the anomalous dimension for the vector operator 
\begin{align}
-\gamma^{a}_{b} = h_{ac}T^{c}_{IJ} g^J h_{bd}T^{d}_{IL} g^{L} \ .
\end{align}
This is essentially Higgs-effect for the background field $a_\mu^a$, which absorbs the gauge directions in the coupling constant space.



The gradient property can be verified by constructing the generating functional explicitly
\begin{align}
S[a_\mu;g] = \int d^d x \sqrt{g} D_\mu g^I D^\mu g^I \ 
\end{align}
with $D_\mu g^I = \partial_\mu g^I +  a^a_\mu h_{ab} T^b_{IJ} g^J$. It is easy to see that the vector beta function is indeed gradient with respect to $S[a_\mu;g]$.
\begin{align}
B^a_\mu = \delta^{ab} g_{\mu\nu} \frac{\delta S[a_\mu;g]}{\delta a^b_\mu} \ .
\end{align}

Finally, we derive the non-renormalization property. When $\hat{\rho}^a_I$ vanishes,  the formula \eqref{vectb} means that the action of the symmetry group on $g^I$ vanishes. Alternatively we recall that \eqref{vectb} is nothing but the one that appears in the right hand side of the conservation equation. Therefore, whenever $\hat{\rho}^a_I$ vanishes the corresponding current must vanish. The converse is also true: when the current is conserved, the corresponding vector beta function vanishes from  \eqref{vectb}.

\subsection{Holography}
One of the basic assumptions of the holography is to identify the gravity partition function (Gubser-Klebanov-Polyakov-Witten partition function) with the Schwinger source functional of the dual field theory. 
In holography, 
the vector beta functions can be interpreted as the radial evolution of the vector fields $A_M^a$ in the bulk $d+1$ dimensional space-time. The structure of the holographic dictionary is better understood when the vector operators in corresponding field theory are conserved. The conserved currents are dual to gauge fields in the bulk. When the gauge symmetry is not spontaneously broken in the bulk, they describe massless vector fields, explaining that the conserved current does not obtain the anomalous dimensions. At the leading order in derivative expansions, the kinetic term is given by the Yang-Mills action.\footnote{In $d=2$, the Chern-Simons interaction can be the leading term.} 

We claim that the spontaneously broken gauge fields in the bulk describe the vector beta functions in the weakly coupled gravity dual.\footnote{A similar idea to study the renormalization group properties of external electric field in the context of holographic realization of condensed matter physics was persued in \cite{Gouteraux:2012yr}\cite{Gouteraux:2013oca}. They also found that what we call the vector beta function is related to the bulk matter current in holography. By restricting a particular form of the source vector fields, they also studied the full back-reaction on the metric, which we will not  discuss.} 
The appearance of the spontaneously broken gauge fields in the bulk may be natural by studying their equations of motion in the bulk
\begin{align}
D^M F^a_{MN} = \mathcal{J}^a_N \ 
\end{align}
in the $\mathrm{AdS}_{d+1}$ background
\begin{align}
ds^2 = g_{MN}^{(d+1)} dx^M dx^N = dr^2 + e^{2Ar} \eta_{\mu\nu} dx^\mu dx^\nu \ .
\end{align}
Here $\mathcal{J}^a_N$ is the matter current charged under the gauge fields $A^a_M$ with the field strength $F^a_{MN}$. Working in $A^a_r = 0$ gauge (more on the gauge choice later), and keeping the slowest fluctuation in $x^\mu$ directions (having power-counting renormalization scheme in mind), the equations of motion become
\begin{align}
\partial^2_r A^a_\mu - (d-2)A \partial_r A^a_\mu = \mathcal{J}^a_\mu \ . \label{eomvector}
\end{align}
Note that, for instance, when the matter action is given by the non-linear sigma model with a potential 
\begin{align}
S_{\mathrm{matter}} = \int d^{d+1} x \sqrt{g^{(d+1)}} \left( G_{IJ}(\Phi) D_M \Phi^I D^M \Phi^J + V(\Phi) \right) \ , \label{scalaraction}
\end{align}
the current $\mathcal{J}^a_\mu$ here is explicitly given by
\begin{align}
\mathcal{J}^a_\mu = \frac{\delta S_{\mathrm{matter}}}{\delta A^a_\mu} = G^{IJ}(\Phi) T^{a I}_{\ \ K}\Phi^K D_\mu \Phi^J \ , \label{current}
\end{align}
where $G^{IJ}(\Phi)$ is gauge invariant metric of the target space.\footnote{The other equation from $A^a_r$ variation is the analogue of Gauss-law, and it is interpreted as the consistency condition of the operator identification under the renormalization group flow.}
When the gauge group is non-Abelian, there are additional contributions from the non-linearity of the gauge kinetic terms, but we will neglect them with the same reason we have discarded the higher $x^\mu$ derivatives from the power-counting argument in the dual field theory side.

In order to connect the second order differential equations to the first order renormalization group equation, we may use the Hamilton-Jacobi method with suitable boundary conditions \cite{de Boer:1999xf}\cite{Papadimitriou:2004ap}\cite{Papadimitriou:2004rz}\cite{Nitti}. Alternatively, for our perturbative purpose, we may use the singular perturbation theory of the differential equations with the renormalization group improvement method \cite{Nakayama:2013fha}.

The idea is to employ the renormalization group technique proposed in \cite{Chen:1994zza} to improve the perturbative solutions of the differential equations such as \eqref{eomvector}. We will encounter the secular terms proportional to $r$ in the naive perturbation theory. To get the better control in large (negative) $r$ region, we had to sum the leading log contributions. This can be done by introducing the fictitious cut-off and demand the initial condition depend on the cut-off but the solution itself is independent of the cut-off. This method of approaching the singular perturbation theory in the bulk is shown to be equivalent to the improved perturbation theory of the dual field theory computation.

Practically, if we would like to obtain the leading order perturbative solutions of the source function $A^a_\mu$ when the condensation of $\Phi^I$ is small near the AdS boundary, we may neglect the higher derivatives in \eqref{eomvector} and solve
\begin{align}
-\frac{1}{A}\frac{\partial A^a_\mu (r)}{ \partial r} = \frac{1}{A^2(d-2)} \mathcal{J}^a_\mu (A_\mu, \Phi^I) \ , \label{maxwell}
\end{align}
where $\Phi^I$ also satisfies the first order renormalization group equation
\begin{align}
\frac{1}{A}\frac{\partial \Phi^I}{ \partial r} = B^I(\Phi) \ ,
\end{align}
which again can be more formally obtained from the Hamilton-Jacobi method or singular perturbation theory mentioned above. In particular, near the massless limit of the non-linear sigma model without considering the back-reaction to the AdS space-time, the current source for the vector fields is given by \eqref{current} and the source for the scalar fields is given by
\begin{align}
B^I = \frac{G^{IJ}\partial_J V(\Phi)}{A^2 d} \ , \label{scalargrad}
\end{align}
where $V(\Phi)$ is gauge invariant (super)potential.\footnote{With the large (negative) cosmological constant, the superpotential and the potential may be identified for the domain wall flow that connects two AdS vacua.}

We will interpret the current $\mathcal{J}^a_\mu(A_\mu,\Phi)$ and the gradient of the potential $B^I(\Phi)$ as the holographic vector and scalar beta functions from the holographic identification of the radial coordinate and the renormalization scale $A r \sim \log\mu$. In the following, we would like to verify the properties of the vector beta functions discussed in section 2 from the holographic construction.

Let us begin with (i) the gauge invariance. The compensated gauge invariance of the $d$ dimensional slice in holography is obviously inherited from the bulk gauge invariance in $d+1$ dimensional space-time. While we have partially fixed the gauge by demanding $A^a_r = 0$, the residual gauge invariance requires that $\mathcal{J}^a_\mu$ is a covariant vector under the $d$-dimensional gauge transformation which does not change $A_r$.\footnote{Similarly, the holographic scalar beta functions transform covariantly as can been seen from \eqref{scalargrad}.} Indeed, the matter action for the non-linear sigma model such as \eqref{scalaraction} and the corresponding current $\mathcal{J}^a_\mu$ in \eqref{current} is precisely the form we have studied in the dual field theory \eqref{vectorbetaff} by identifying $\Phi^I$ with $g^I$ which is valid within the weakly coupled holographic scheme of the renormalization group: 
\begin{align}
\hat{\rho}^a_I(\Phi) = \frac{1}{A^2(d-2)} G_{IJ}(\Phi) T^{aJ}_{\ \ K} \Phi^K \ . \label{veccc}
\end{align}

Let us now perform the $r$ dependent gauge transformation with the gauge parameter $\Lambda^a(r,x^\mu)$. Denoting $A^{-1}\partial_r \Lambda^a = v^a$, the ``beta functions" (i.e. radial evolution of the bulk fields) are modified as 
\begin{align}
\delta \beta^I &= v^a T^{aI}_{\ \ J} \Phi^J \cr
\delta \beta_\mu^a &= D_\mu v^a = \partial_I v^a D_\mu \Phi^I \ .
\end{align}
This is equivalent to Class 1 ambiguity the of beta functions \eqref{Bfunction} when $v^a$ does not explicitly depend on $r$. Generically, we may allow an arbitrary dependence on $r$ for $v^a$. We could have considered the compensated gauge transformation that depends on the renormalization scale explicitly in the field theory side while it is not standard.
Furthermore, we may interpret $A^a_r$ as the virial current contribution 
\begin{align}
\delta A_r^a = v^a \ .
\end{align}
This identification was crucial in constructing the holographic models for scale invariant but non-conformal systems studied in \cite{Nakayama:2009qu}\cite{Nakayama:2009fe}\cite{Nakayama:2010wx}.\footnote{If the AdS background with $A_r^a = \mathrm{const}$ and $\Phi^I = \mathrm{const}$ solved the bulk equations of motion, it would mean the dual theory is scale invariant, but not conformal invariant.} As is clear from the discussion, the covariant derivative in the radial direction corresponds to the scheme independent $B^I$ functions, while the ordinary derivative in the radial direction corresponds to the scheme dependent $\beta^I$ functions.

Having established the compensated gauge invariance, the orthogonality can be understood as the gauge invariance of the (super)potential function: 
\begin{align}
0 = \delta^a \Phi^I \partial_I V(\Phi) = T^{aI}_{\ \ K} \Phi^K \partial_I V(\Phi) \ . \label{gaugeinv}
\end{align}
From the explicit form of the holographic scalar beta functions \eqref{scalargrad}, and the vector beta functions \eqref{veccc}, we can infer
\begin{align}
\hat{\rho}_I^a B^I = 0 \ 
\end{align}
from the gauge invariance of the (super)potential and the fact that the same $G_{IJ}$ appears both in $B^I$ and $\hat{\rho}_I^a$. More abstractly, the orthogonality was a consequence of the consistency of the renormalization group operator $\Delta_{\sigma}$. In the holographic side, one may state it as a consequence of the consistency of the radial Hamiltonian evolution. This is guaranteed by the matter equations of motion and the gauge invariance of the action.

The computation of the anomalous dimension matrix in holography is precisely related the Higgs mechanism in the bulk. The vacuum expectation value of $\Phi^I$ we have assumed to spontaneously break the conservation of the current induces the mass for the gauge field $A_\mu$ in the bulk. From the explicit formula \eqref{maxwell} and \eqref{veccc}, we see that the anomalous dimension for $A_\mu$ can be computed by the leading radial evolution, and 
\begin{align}
-\gamma^a_b = \frac{1}{A^2(d-2)}G_{IJ}(\Phi) T^{aI}_{\ \ K} \Phi^K T^{bJ}_{\ \ L} \Phi^L \ 
\end{align}
as expected from the field theory Callan-Symanzik equation \eqref{anovect}. Again, this result is equivalent to the holographic renormalizability of the Gubser-Klebanov-Polyakov-Witten partition function.

The gradient property is a direct consequence of the definition of the current operator in the bulk. By identifying $\Phi^I$ with $g^I$ and $A^a_\mu$ with $a^a_\mu$, the gauged kinetic term of the bulk scalar field $\Phi^I$ in the ``radial Lagrangian" is
\begin{align}
L = \frac{1}{A^2(d-2)}\int d^d x \sqrt{g} G_{IJ} D_\mu \Phi^I D^\mu \Phi^J
\end{align}
and it serves as the potential functional of the vector beta function
\begin{align}
B^a_\mu = g_{\mu\nu} \frac{\delta L}{\delta A^a_\nu} \ .
\end{align}
It is interesting to observe that the potential function for the scalar beta function is essentially given by the (super)potential of the bulk scalar action while the potential functional for the vector beta function is its kinetic term.

Finally we discuss the non-renormalization property. From the Higgs-mechanism discussed above, we can see that whenever the gauge invariance is broken in the bulk by the Higgs-mechanism, the gauge field acquires the anomalous dimension as long as the kinetic terms for the scalars are not degenerate. The degenerate kinetic terms  indicate the breakdown of the null energy-condition, which leads to the violation of unitarity or causality.
On the other hand, the zero anomalous dimension requires the massless gauge field in the bulk, which means that there exists the corresponding conserved current in the dual field theories from the necessity of gauge symmetry for the massless fields from unitarity.
Note that the unitarity seems crucial. One may find counter-examples once we discard the null energy-condition  \cite{Nakayama:2009qu}\cite{Nakayama:2009fe}\cite{Nakayama:2010wx} or the full space-time diffeomorphism \cite{Nakayama:2012sn} of the bulk theory.

\section{Discussions}

In this paper, we have proposed various properties of vector beta functions that govern the renormalization of vector operators in relativistic field theories.  Many of the proposed properties can be obtained by demanding the consistency of the local renormalization group within the power-counting renormalization scheme. We have seen that the existence of such non-trivial properties seem very natural in holographic renormalization group and probably they are also necessary for the consistency of the bulk theories with unitarity and causality. Indeed some of the properties are tightly connected with the $d+1$ dimensional gauge invariance and the null energy condition.

For the scalar beta functions, one important observation that follows from the gradient property is the so-called ``$c$-theorem" that claims there exists a monotonically decreasing function along the renormalization group flow. The weak version of the $c$-theorem that compares the values at conformal fixed points have been proven to be true in even dimensions \cite{Zamolodchikov:1986gt}\cite{Komargodski:2011vj} beyond the power-counting renormalization under some technical assumptions such as unitarity. Physical intuition behind is that we lose information along the renormalization group flow by course graining, so some sorts of ``entropy" must decrease. However, this intuitive argument does not explain why in non-unitary or non-relativistic field theories, cyclic renormalization group flow seems possible.

A natural question we should ask is if there is a generalization when the ``coupling constants" are space-time dependent and the vector sources are non-zero. Naively, one may try to simply forget the contributions from the vector source and define the local ``$c$-function" out of space-time dependent coupling constant. It is true that the so-constructed local  ``$c$-function", at least within the power-counting renormalization scheme, generates the gradient flow for the scalar beta functions, but since local renormalization group flow should contain the information of the vector source, it is not satisfactory.\footnote{From the local Callan-Symanzik equation, we see that one cannot trade the derivative with respect to the local renormalization scale only with the derivatives with respect to the local coupling constants. It needs the additional contributions from the change of the vector source.}

We could imagine that the potential functional $S_{\mathrm{vector}}[a_\mu,g^I]$ for the vector beta functions may play a similar role to the potential function $\hat{a}(g^I)$ for the scalar beta functions, showing the generalized monotonicity. Assuming $H^{ab}$ is positive definite, it may work in the Euclidean signature. However, since $B_\mu^a$ can be in either time-direction or space-directions, the overall signature can easily change by itself. Nevertheless, one may still expect that if we add the contributions from the scalar beta functions $S_{\mathrm{scalar}}[g^I] = \int d^d x\sqrt{g} \hat{a}(g^I)$, then the full master functional $S_{\mathrm{master}}[a_\mu,g^I] = S_{\mathrm{vector}}[a_\mu,g^I] + S_{\mathrm{scalar}}[g^I] $ may show a better behavior under the combined renormalization group flow with respect $g^I(x)$ and $a_\mu(x)$. In the holographic dual, it is this combined matter Lagrangian that can show the local positivity from the null energy condition.

We may still wonder how to interpret the variation of $S_{\mathrm{master}}[a_\mu,g^I]$ with respect to $g^I$. We have understood that the contribution from the ``potential term" generates the scalar beta functions.
The contribution from the ``kinetic term", however, gives the second order kinetic energy of $g^I$ which cannot appear in the scalar beta functions in the power-counting renormalization. Technically speaking, this is one point that our formulation does not agree with the holographic formulation proposed in \cite{Lee:2013dln}, where the gradient properties are assumed for the master functional $S_{\mathrm{master}}[a_\mu,g^I]$ including every tensor operators. What seems to happen in holography in Hamiltonian-Jacobi formulation is that the second derivative terms are removed by the boundary condition. Since it is not obvious how to relate the second derivatives of source functional to the lower order derivatives by using ``equations of motion of the source", the role of the master functional remains open without holography.\footnote{On the other hand, for relevant operators with dimension $d-2$, there do exist the second derivative contributions to the renormalization of the coupling constants \eqref{mderiv}. We note, however, in this case there is no contributions to the vector beta functions.}

Apart from holography, one interesting theoretical application of vector beta functions is the various domain walls in quantum field theories. So-called duality walls or renormalization group walls \cite{Gaiotto:2012np} introduce the space-time dependent coupling constant which smoothly or possibly discontinuously varies at or near the wall. Most of domain walls so-far studied in the literature have coupling constants that are not charged under the (partially broken) symmetry of the theory, but one can imagine that in most generality, they should induce the renormalization of the vector sources through the vector beta functions. It should be very important to understand the general structure of the renormalization of the domain walls based on the properties proposed in this paper.

Another interesting application would be the new renormalization group fixed point associated with the space-time dependent coupling constant advocated in \cite{Dong:2012ua}. They did not study the induced vector operators that may appear once the scalar coupling constants are space-time dependent. One remark is that their action of dilatation is slightly different from ours because they fix the origin of the space-time and act the dilatation directly on coordinate $x^\mu$ rather than coordinate difference $dx^\mu$ employed here. Regardless of this remark, it would be important to understand the effects of the vector beta functions there to understand the nature of these new fixed points.

Finally, we conclude by mentioning that there are at least 72 vector beta functions that should be computed in the standard model of our universe.

\section*{Acknowledgements}
This work is supported by the World Premier International Research
Center Initiative (WPI Initiative), MEXT, Japan.


\begin{thebibliography}{99}
\bibitem{Osborn:1991gm} 
  H.~Osborn,
  Nucl.\ Phys.\ B {\bf 363}, 486 (1991).

\bibitem{JO}
I.~Jack and H.~Osborn. 
To appear.



\bibitem{Lee:2012xba} 
  S.~-S.~Lee,
  JHEP {\bf 1210}, 160 (2012)
  [arXiv:1204.1780 [hep-th]].
\bibitem{Lee:2013dln} 
  S.~-S.~Lee,
  arXiv:1305.3908 [hep-th].



\bibitem{Nakayama:2013wda} 
  Y.~Nakayama,
  arXiv:1307.8048 [hep-th].


\bibitem{Jack:1990eb} 
  I.~Jack and H.~Osborn,
  Nucl.\ Phys.\ B {\bf 343}, 647 (1990).

\bibitem{Nakayama:2013is} 
  Y.~Nakayama,
  arXiv:1302.0884 [hep-th].



\bibitem{Nakayama:2012sn} 
  Y.~Nakayama,
  Gen.\ Rel.\ Grav.\  {\bf 44}, 2873 (2012)
  [arXiv:1203.1068 [hep-th]].

\bibitem{Grinstein:2013cka} 
  B.~Grinstein, A.~Stergiou and D.~Stone,
  arXiv:1308.1096 [hep-th].



\bibitem{Wallace:1974dy} 
  D.~J.~Wallace and R.~K.~P.~Zia,
  Annals Phys.\  {\bf 92}, 142 (1975).

\bibitem{Zamolodchikov:1986gt} 
  A.~B.~Zamolodchikov,
  JETP Lett.\  {\bf 43}, 730 (1986)
  [Pisma Zh.\ Eksp.\ Teor.\ Fiz.\  {\bf 43}, 565 (1986)].

\bibitem{Komargodski:2011vj} 
  Z.~Komargodski and A.~Schwimmer,
  JHEP {\bf 1112}, 099 (2011)
  [arXiv:1107.3987 [hep-th]].

\bibitem{Mack:1975je} 
  G.~Mack,
  Commun.\ Math.\ Phys.\  {\bf 55}, 1 (1977).



\bibitem{Nakayama:2012nd} 
  Y.~Nakayama,
  Phys.\ Rev.\ D {\bf 87}, 085005 (2013)
  [arXiv:1208.4726 [hep-th]].

\bibitem{Luty:2012ww} 
  M.~A.~Luty, J.~Polchinski and R.~Rattazzi,
  JHEP {\bf 1301}, 152 (2013)
  [arXiv:1204.5221 [hep-th]].

\bibitem{Fortin:2012hn} 
  J.~-F.~Fortin, B.~Grinstein and A.~Stergiou,
  JHEP {\bf 1301}, 184 (2013)
  [arXiv:1208.3674 [hep-th]].


\bibitem{Dymarsky:2013pqa} 
  A.~Dymarsky, Z.~Komargodski, A.~Schwimmer and S.~Theisen,
  arXiv:1309.2921 [hep-th].

\bibitem{Farnsworth:2013osa} 
  K.~Farnsworth, M.~A.~Luty and V.~Prelipina,
  arXiv:1309.4095 [hep-th].

\bibitem{Jack:1999aj} 
  I.~Jack and D.~R.~T.~Jones,
  Phys.\ Lett.\ B {\bf 465}, 148 (1999)
  [hep-ph/9907255].

\bibitem{Fortin:2012hc} 
  J.~-F.~Fortin, B.~Grinstein, C.~W.~Murphy and A.~Stergiou,
  Phys.\ Lett.\ B {\bf 719}, 170 (2013)
  [arXiv:1210.2718 [hep-th]].



\bibitem{Friedan:2009ik} 
  D.~Friedan and A.~Konechny,
  J.\ Phys.\ A {\bf 43}, 215401 (2010)
  [arXiv:0910.3109 [hep-th]].


\bibitem{Gouteraux:2012yr} 
  B.~Gouteraux and E.~Kiritsis,
  JHEP {\bf 1304}, 053 (2013)
  [arXiv:1212.2625 [hep-th]].

\bibitem{Gouteraux:2013oca} 
  B.~Gouteraux,
  arXiv:1308.2084 [hep-th].

\bibitem{de Boer:1999xf} 
  J.~de Boer, E.~P.~Verlinde and H.~L.~Verlinde,
  JHEP {\bf 0008}, 003 (2000)
  [hep-th/9912012].

\bibitem{Papadimitriou:2004ap} 
  I.~Papadimitriou and K.~Skenderis,
  hep-th/0404176.


\bibitem{Papadimitriou:2004rz} 
  I.~Papadimitriou and K.~Skenderis,
  JHEP {\bf 0410}, 075 (2004)
  [hep-th/0407071].

\bibitem{Nitti}
E.~Kiritsis, F.~Nitti and W.~Li,
to appear.

\bibitem{Nakayama:2013fha} 
  Y.~Nakayama,
  arXiv:1305.4117 [hep-th].



\bibitem{Chen:1994zza} 
  L.~Y.~Chen, N.~Goldenfeld and Y.~Oono,
  Phys.\ Rev.\ Lett.\  {\bf 73}, 1311 (1994).

\bibitem{Nakayama:2009qu} 
  Y.~Nakayama,
  JHEP {\bf 0911}, 061 (2009)
  [arXiv:0907.0227 [hep-th]].

\bibitem{Nakayama:2009fe} 
  Y.~Nakayama,
  JHEP {\bf 1001}, 030 (2010)
  [arXiv:0909.4297 [hep-th]].

\bibitem{Nakayama:2010wx} 
  Y.~Nakayama,
  Eur.\ Phys.\ J.\ C {\bf 72}, 1870 (2012)
  [arXiv:1009.0491 [hep-th]].


\bibitem{Nakayama:2012sn} 
  Y.~Nakayama,
  Gen.\ Rel.\ Grav.\  {\bf 44}, 2873 (2012)
  [arXiv:1203.1068 [hep-th]].

\bibitem{Gaiotto:2012np} 
  D.~Gaiotto,
  JHEP {\bf 1212}, 103 (2012)
  [arXiv:1201.0767 [hep-th]].

\bibitem{Dong:2012ua} 
  X.~Dong, B.~Horn, E.~Silverstein and G.~Torroba,
  Phys.\ Rev.\ D {\bf 86}, 105028 (2012)
  [arXiv:1207.6663 [hep-th]].

\end{thebibliography}
\end{document}